\documentclass[11pt]{article}
\usepackage{setspace}
\usepackage{amssymb}
\usepackage{amsmath}
\usepackage{amsfonts}
\usepackage{slashed}

\def\be{\begin{equation}}

\def\ee{\end{equation}}

\def\bea{\begin{eqnarray}}

\def\eea{\end{eqnarray}}

\newcommand\eps{\epsilon}

\makeatletter
\def\blfootnote{\xdef\@thefnmark{}\@footnotetext}
\makeatother

\begin{document}

\singlespace

\begin{flushright} BRX TH-6656 \\
CALT-TH 2019-37
\end{flushright}

\vspace*{.3in}

\begin{center}

{\Large\bf Why does $D=4$, rather than more (or less)? An Orwellian explanation}

{\large S.\ Deser}

{\it 
Walter Burke Institute for Theoretical Physics, \\
California Institute of Technology, Pasadena, CA 91125; \\
Physics Department,  Brandeis University, Waltham, MA 02454 \\
{\tt deser@brandeis.edu}
}
\end{center}

\begin{abstract}
Two simple, if draconian, assumptions about classical gravity actions fix space-time's dimension at D=4. 

\end{abstract}
For Mike Duff, friend, collaborator, eminent theoretician, peerless Defender of the Faith, on his 70th. 

\section{Introduction}
Einstein's greatest achievement was to liberate space-time from its fixed, a priori, flat background status to become a dynamical system on a par with the rest of the Universe, namely a pseudo-Riemann manifold determined by its own field equations. The pendant to this unique insight would be a dynamic determination of its dimensionality as well, a more ``discrete" challenge, yet to be met. Of course, we have learnt through the work of Ken Wilson and others that reality consists of layers, as we descend in distance or go up in energy, and that classical GR, and with it the very notion of dimension, is bound to be profoundly altered, if not completely discarded, in future. But effective actions have been so powerful that we should try to maximize their predictions; even if $D$ is ``really" $5$, $10$,$11$, some other $4+n$, or none of the above, it would be instructive to see why $4$ is a pretty good physical choice at our level.

Some time ago, I surveyed the ``Many dimensions of dimension" [1] in honor of Gunnar Nordstrom, who first proposed a higher, $D=5$, dimensional world, a century ago; he must have felt how daring his proposal was! Nowadays, we hardly blink at a variety of much higher $D$, especially in quantum models, without a thought for whether $D=4$ can at least be reasonably justified on its own grounds. The present attempt presents an argument from classical gravity that $D=4$, on the basis of two simple, if possibly stringent, assumptions.

\section{The assumptions}
 \begin{enumerate}
\item What is not forbidden is compulsory: the gravitational action in any dimension $D$ consists of ALL local curvature invariants and their derivatives, with non-zero coefficients, that result in second order field equations. The latter requirement is of course a sacred classical Newtonian one, that needs no justification. 
\item Only models with one (stable) constant curvature vacuum are acceptable.
\end{enumerate}
To amplify, the Orwellian aspect of A is that one must keep all legal terms in the action, rather than pre-discard what we don't like, as is normally done; indeed, quantum corrections (say) are likely to bring them all in. The hidden purpose of B is to get our desired dimension; however, it is a reasonable physical requirement on its own. Fine print points will be dealt with as they arise.

\section{The framework}

As stated, we assume gravity to be a dynamical pseudo-Riemannian manifold (signature determination is not yet in our ambit!) of dimension $D$, subject to two reasonable (to the author, at least) assumptions. The first, hardly arguable, one is that the action be local and produce second derivative order field equations (indeed that's how GR can be uniquely defined in $D=4$), but more strongly that ALL possible such terms must be included {\it a priori}. This latter assumption is also reasonable, not yet having made any choice of basic dynamics, such as GR. The first fine print point here is that this (only) seemingly excludes all higher curvature corrections, but one must differentiate as usual between the basic action and its ``external" corrections, that are to be treated only as vertex, rather than ``propagator", contributions, in the sense of effective theory. We will discuss the, essentially irrelevant, cosmological term below.

Let us get the lower $D$ out of the way first. The $D=2$ Einstein action is a topological invariant and produces no field equations, while all other Lagrangians, such as $R^2$, give higher order ones. For $D=3$, (only) the Einstein term is second order but flat space is its only solution, so back to rigid backgrounds, while again all other choices raise the derivative order.\footnote{Topologically massive gravity [2] is an interesting borderline case. While manifestly of third derivative order, hence excluded by A, it is effectively, if nontrivially, second order and possesses rich dynamics in D=3.}
At exactly $D=4$, only the desired Einstein action is allowed a priori, There remains the task of excluding $D>4$. [The so-called critical point gravities in $D>3$ [5] are excluded for having multiple vacua.] There, in addition to the Einstein action, a tower increasing with D, of candidates respecting A arises: the Lovelock terms [3]. They are based on the Bianchi identity $D_{[\mu} R_{\nu\rho]\alpha\sigma}=0$, and have the generic form\footnote{Parenthetically, it has been shown [4] that generic Lovelock gravity has $m=0$, $s=2$ excitations about its constant curvature vacua, but this is of course irrelevant to our Orwellian context.}
\begin{equation}
I_L = \int d^D x \eps^{\alpha \mu \beta\rho \ldots} \eps^{\gamma\tau\delta\sigma \ldots } R_{\alpha \beta \gamma \delta} \ldots R_{\mu\rho \tau\sigma} g_{..} g_{..}/\sqrt{-g}
\end{equation}
%I_L= ?d^D x eps^? eps^xxx.. R_ab cd?R_mp qr g_..g_../?-g.      (1)
By Bianchi, variations of the curvatures do not contribute, but only those of the metric/vielbein, so the field equations only contain the second derivative undifferentiated curvatures. These terms obviously require $D>4$: at $D=4$, the epsilons' $2\times 4= 8$ indices are saturated by two $R$, leaving no room for metrics to be varied; indeed (1) reduces to the Gauss-Bonnet invariant  $\int d^4 x(\hbox{Riem}^2 - 4 \hbox{Ricc}^2 + R^2)$.
Note that there are two types of towers as $D$ increases, those with a fixed number ($2$ or greater) of Riemanns and those with a rising number of them, at the expense of fewer metrics. We are pledged to keep both types of towers, also those with more than two $R$s.

\section{Vacua}

We turn now to the role of our second requirement, B, that acceptable models have only one (stable) vacuum. Here all the work has been done for us in [6] (amplified in [7]), for a two-term action such as arises in $D=5,6$ when only the $R^2$ Lovelock term is present. The point of [6] was that there are necessarily two vacuum solutions to the sum of Einstein and Lovelock actions, namely flat space and (A)dS; in the latter, neither part vanishes but the two cancel each other, independent of the presence of a bare cosmological term. While both solution are generically stable [7], they violate our injunction B, that only one vacuum be allowed. So we must, in obeisance to A, reject ALL $D>4$ models since all are at least two-vacuum. At every $D+2$ dimension, a new higher $R^{n+1} \rightarrow \Lambda^{n+1}$ term becomes available.
The most general vacua at $D$ are then the constant (or $0$) curvature roots of the algebraic equation
\begin{equation}
\sum_{n=1}^{\lfloor (D-1)/2 \rfloor} a_n \Lambda^n = 0,
\end{equation}
%a_0 /\_0 +CAP SIGMA_n=1 ^ [D/2] a_n /\^n=0       (2)
where $\lfloor (D-1)/2 \rfloor$ is the integral part of $(D-1)/2$; by A, a possible cosmological $a_1 \Lambda^1$ term must be included, since its contribution to the field equations has no derivatives.  Then there are manifestly at least two vacuum branches for $D>4$, while for $D=3$, $4$, where Lovelock terms are absent, only the single vacuum $a_1 \Lambda^1$ remains, as discussed above. [At $D=2$, keeping a non-zero $\Lambda$ would yield the field equation $g_{\mu\nu} = 0$, while dropping it yields no equation at all.]  To be quite clear, irrespective of dimension, there will always be multiple solutions of (2) if all $a_n \ne  0$ per Orwell.  The powers of curvature, that is of $\Lambda$, from Lovelock are limited to be $>1$ and $< \lfloor (D+1)/2\rfloor$, there is only the (Einstein) power $R^1 \sim \Lambda^1$ in any $D$ from the non-Lovelock part (by A). Since the sum (2) starts as $\Lambda^1$, flat space is always a solution, as is at least one $\Lambda \ne 0$. Indeed, any multiple term action necessarily has at least flat space as well as one other solution.  In $D=4$, exceptionally, there are just two separate $\Lambda^1$ terms, so either we choose flat space, $\Lambda=0$, or must tune the two a to cosmological GR. 
The presence of normal, positive, matter sources does not affect the above reasoning, based on the left hand sides of gravity, except that of course they must be $D=4$-compatible.

\section{Discussion}
Our two conditions forcing $D=4$, that each action term at given $D$ yielding second order equations is to be included {\it en bloc}, and that vacuum is a singular noun, may perhaps be too Draconian. Both are needed: lifting either allows for any $D$. Still, they do provide a sufficient condition for $D=4$ already at classical gravitational level as Einstein would suggest, a result not previously achieved.

\section*{Acknowledgement}
The work of SD was supported by the U.S. Department of Energy, Office of Science, Office of High Energy Physics, under Award Number de-sc0011632; I thank Bayram Tekin for useful correspondence and J Franklin for great Tex help.

\end{document}